
\magnification \magstep1
\raggedbottom
\openup 4\jot
\voffset6truemm
\headline={\ifnum\pageno=1\hfill\else
\hfill{\it Boundary terms in complex general relativity}
\hfill \fi}
\rightline {June 1995, DSF preprint 95/25}
\centerline {\bf BOUNDARY TERMS IN COMPLEX GENERAL RELATIVITY}
\vskip 1cm
\centerline {\bf Giampiero Esposito and Cosimo Stornaiolo}
\vskip 0.3cm
\noindent
{\it Istituto Nazionale di Fisica Nucleare,
Sezione di Napoli, Mostra d'Oltremare Padiglione 20,
80125 Napoli, Italy;}
\vskip 0.3cm
\noindent
{\it Dipartimento di Scienze Fisiche, Mostra d'Oltremare
Padiglione 19, 80125 Napoli, Italy.}
\vskip 0.3cm
\noindent
{\bf Abstract}. A recent analysis of
real general relativity based on multisymplectic techniques
has shown that boundary terms may occur in the constraint
equations, unless some boundary conditions are imposed.
This paper studies the corresponding form of such boundary
terms in complex general relativity, where space-time is
a four-complex-dimensional complex-Riemannian
manifold. A complex Ricci-flat
space-time is recovered providing some boundary conditions
are imposed on two-complex-dimensional surfaces.
One then finds that the holomorphic multimomenta should
vanish on an arbitrary three-complex-dimensional surface,
to avoid having
restrictions at this surface on the spinor fields
which express the invariance of the theory under holomorphic
coordinate transformations. The Hamiltonian constraint
of real general relativity is then replaced by a geometric
structure linear in the holomorphic multimomenta, and a link
with twistor theory is found. Moreover, a deep relation
emerges between complex space-times which are not
anti-self-dual and two-complex-dimensional surfaces
which are not totally null.
\vskip 100cm
\noindent
\leftline {\bf 1. Introduction}
\vskip 1cm
\noindent
Among the various approaches to the quantization of the
gravitational field, much insight has been gained by the
use of twistor theory and Hamiltonian techniques (see [1-5]
and references therein). For example, it is by now well-known
how to reconstruct an anti-self-dual space-time out of
deformations of flat projective twistor space, and the various
definitions of twistors in curved space-time enable one to
obtain relevant information about complex space-time geometry
within a holomorphic, conformally invariant framework.
Moreover, the recent approaches to canonical gravity described
in [3] have led to many exact solutions of the quantum
constraint equations of general relativity, although their
physical relevance for the quantization programme remains
unclear. A basic difference between the Penrose formalism
[1-2,5] and the Ashtekar formalism [3] is as follows.
The twistor programme refers to a four-complex-dimensional
complex-Riemannian manifold with holomorphic metric,
holomorphic connection and holomorphic curvature tensor,
where the complex Einstein equations are imposed. By contrast,
in the recent approaches to canonical gravity, one studies
complex tetrads on a four-real-dimensional Lorentzian
manifold, and real general relativity may be recovered
providing one is able to impose suitable reality conditions.

The aim of this paper is to describe a new property of
complex general relativity within the holomorphic framework
relevant for twistor theory, whose derivation results from
recent attempts to obtain a manifestly covariant formulation
of Ashtekar's programme [6]. For this purpose, section 2
studies boundary conditions relevant for the multisymplectic
description of Lorentzian space-times, whilst their
holomorphic counterpart appears in section 3. The multisymplectic
form of complex general relativity, with the corresponding
equations, which are linear in the holomorphic multimomenta,
is studied in section 4. Open problems are presented in
section 5.
\vskip 1cm
\leftline {\bf 2. Boundary conditions in the Lorentzian theory}
\vskip 1cm
\noindent
It has been recently shown in [6] that the constraint
analysis of general relativity may be performed by using
multisymplectic techniques, without relying on a 3+1 split
of the space-time four-geometry. The constraint equations
(cf section 4)
have been derived while paying attention to boundary terms, and
the Hamiltonian constraint turns out to be linear in the
{\it multimomenta}. Whilst the latter property is more
relevant for the (as yet unknown) quantum theory of gravitation,
the former result on boundary terms deserves further thinking
already at the classical level, and is the object of our
investigation.

We here write the Lorentzian space-time 4-metric as
$$
g_{ab}=e_{a}^{\; \; {\hat a}} \; e_{b}^{\; \; {\hat b}}
\; \eta_{{\hat a}{\hat b}}
\eqno (2.1)
$$
where $e_{a}^{\; \; {\hat a}}$ is the cotetrad and $\eta$ is
the Minkowski metric. In first-order theory, the tetrad
$e_{a}^{\; \; {\hat a}}$ and the connection 1-form
$\omega_{a}^{\; \; {\hat b}{\hat c}}$ are regarded as
independent variables. In [6] it has been shown that, on
using jet-bundle formalism and covariant multimomentum
maps (see appendix),
the constraint equations of real general relativity
hold on an {\it arbitrary} three-real-dimensional
hypersurface $\Sigma$ providing one of the following
three conditions holds:
\vskip 0.3cm
\noindent
(i) $\Sigma$ has no boundary;
\vskip 0.3cm
\noindent
(ii) the multimomenta
${\tilde p}_{\; \; \; {\hat c}{\hat d}}^{ab}
\equiv e \Bigr(e_{\; \; {\hat c}}^{a} \;
e_{\; \; {\hat d}}^{b}
-e_{\; \; {\hat c}}^{b} \; e_{\; \; {\hat d}}^{a}
\Bigr)$ vanish at $\partial \Sigma$, $e$ being the
determinant of the tetrad;
\vskip 0.3cm
\noindent
(iii) an element of the algebra $o(3,1)$
corresponding to the gauge group, represented by
the antisymmetric $\lambda^{{\hat a}{\hat b}}$,
vanishes at $\partial \Sigma$, and the connection 1-form
$\omega_{a}^{\; \; {\hat b}{\hat c}}$ or $\xi^{b}$ vanishes
at $\partial \Sigma$, $\xi$ being a vector field describing
diffeomorphisms on the base-space.
\vskip 0.3cm
\noindent
In other words, boundary terms may occur in the constraint
equations of real general relativity, and they result from
the total divergences of [6]
$$
\sigma^{ab} \equiv {\tilde p}_{\; \; \; {\hat c}{\hat d}}^{ab}
\; \lambda^{{\hat c}{\hat d}}
\eqno (2.2)
$$
$$
\rho^{ab} \equiv {\tilde p}_{\; \; \; {\hat c}{\hat d}}^{ab}
\; \omega_{f}^{\; \; {\hat c}{\hat d}}
\; \xi^{f}
\eqno (2.3)
$$
integrated over $\Sigma$.

In two-component spinor language, denoting by
$\tau_{\; \; BB'}^{{\hat a}}$ the Infeld-van der Waerden
symbols, the two-spinor version of the tetrad reads
$$
e_{\; \; BB'}^{a} \equiv e_{\; \; {\hat a}}^{a} \;
\tau_{\; \; BB'}^{{\hat a}}
\eqno (2.4)
$$
which implies that $\sigma^{ab}$ in (2.2) takes the form
$$
\sigma^{ab}=e \Bigr(e_{\; \; CC'}^{a} \;
e_{\; \; DD'}^{b}-e_{\; \; DD'}^{a} \;
e_{\; \; CC'}^{b}\Bigr)
\tau_{{\hat a}}^{\; \; CC'} \;
\tau_{{\hat b}}^{\; \; DD'}
\; \lambda^{{\hat a}{\hat b}}.
\eqno (2.5)
$$
Thus, on defining the spinor field
$$
\lambda^{CC'DD'} \equiv \tau_{{\hat a}}^{\; \; CC'}
\; \tau_{{\hat b}}^{DD'} \; \lambda^{{\hat a}{\hat b}}
\equiv \Lambda_{1}^{(CD)} \; \epsilon^{C'D'}
+\Lambda_{2}^{(C'D')} \; \epsilon^{CD}
\eqno (2.6)
$$
the first of the boundary conditions in (iii) is satisfied
providing $\Lambda_{1}^{(CD)}=0$ at $\partial \Sigma$ in real
general relativity, since then $\Lambda_{2}^{(C'D')}$ is obtained
by complex conjugation of $\Lambda_{1}^{(CD)}$, and hence the
condition $\Lambda_{2}^{(C'D')}=0$ at $\partial \Sigma$ leads to
no further information.
\vskip 1cm
\leftline {\bf 3. Boundary conditions in the holomorphic framework}
\vskip 1cm
\noindent
In the {\it holomorphic} framework, no complex conjugation
relating primed to unprimed spin-space can be defined, since
such a map is not invariant under holomorphic coordinate
transformations [1,5]. Hence spinor fields belonging to
unprimed or primed spin-space are {\it totally independent},
and the first of the boundary conditions in (iii) reads
$$
\Lambda^{(CD)}=0
\; \; \; \; {\rm at} \; \partial \Sigma_{c}
\eqno (3.1)
$$
$$
{\widetilde \Lambda}^{(C'D')}=0
\; \; \; \; {\rm at} \; \partial \Sigma_{c}
\eqno (3.2)
$$
where $\partial \Sigma_{c}$ is a two-complex-dimensional
complex surface, bounding the three-complex-dimensional
surface $\Sigma_{c}$, and the {\it tilde} is used to
denote {\it independent} spinor fields [1,5], not related
by any conjugation.

Similarly, $\rho^{ab}$ in (2.3) takes the form
$$
\rho^{ab}=e \Bigr(e_{\; \; CC'}^{a} \;
e_{\; \; DD'}^{b}
-e_{\; \; DD'}^{a} \; e_{\; \; CC'}^{b} \Bigr)
\Bigr(\Omega_{f}^{(CD)} \; \epsilon^{C'D'}
+{\widetilde \Omega}_{f}^{(C'D')} \;
\epsilon^{CD}\Bigr)\xi^{f}
\eqno (3.3)
$$
and hence the second of the boundary conditions in (iii)
leads to the {\it independent} boundary conditions
$$
\Omega_{f}^{(CD)}=0
\; \; \; \; {\rm at} \; \partial \Sigma_{c}
\eqno (3.4)
$$
$$
{\widetilde \Omega}_{f}^{(C'D')}=0
\; \; \; \; {\rm at} \; \partial \Sigma_{c}
\eqno (3.5)
$$
in complex general relativity. The equations (3.4)-(3.5) may be
replaced by the condition
$$
u^{AA'}=0 \; \; \; \; {\rm at} \; \partial \Sigma_{c}
\eqno (3.6)
$$
where $u$ is a holomorphic vector field describing holomorphic
coordinate transformations on the base-space, i.e. on
complex space-time.
\vskip 1cm
\leftline {\bf 4. Multisymplectic form of complex general relativity}
\vskip 1cm
\noindent
The picture of complex general relativity resulting from sections
2-3, and from the analysis in [6], is
highly non-trivial. One starts from a one-jet bundle $J^{1}$
which, in local coordinates, is described by a holomorphic
coordinate system, with holomorphic tetrad, holomorphic
connection 1-form $\omega_{a}^{\; \; {\hat b}{\hat c}}$,
multivelocities corresponding to the tetrad and
multivelocities corresponding to
$\omega_{a}^{\; \; {\hat b}{\hat c}}$, both of holomorphic
nature. The intrinsic form of the field equations, which is a
generalization of a mathematical structure already existing in
classical mechanics, leads to the complex vacuum Einstein
equations $R_{ab}=0$, and to a condition on the covariant
divergence of the multimomenta. Moreover, the covariant
multimomentum map (see appendix and [6]),
evaluated on a section of $J^{1}$ and integrated on an
arbitrary three-complex-dimensional surface $\Sigma_{c}$,
reflects the
invariance of complex general relativity under all holomorphic
coordinate transformations. Since space-time is now a complex
manifold, one deals with holomorphic coordinates which are
all on the same footing, and hence no time coordinate can be
defined. Thus, the {\it constraints} result from the holomorphic
version of the covariant multimomentum map, but cannot be related
to a Cauchy problem as in the Lorentzian theory (cf [7] and
references therein). In particular, the Hamiltonian constraint
of Lorentzian general relativity is replaced by a geometric
structure which
is linear in the holomorphic multimomenta, providing two
boundary terms can be set to
zero (of course, our multimomenta are holomorphic by
construction, since in complex general relativity the
tetrad is holomorphic).
For this purpose, one of the following three conditions
should hold:
\vskip 0.3cm
\noindent
(i) $\Sigma_{c}$ has no boundary;
\vskip 0.3cm
\noindent
(ii) the holomorphic multimomenta vanish at
$\partial \Sigma_{c}$;
\vskip 0.3cm
\noindent
(iii) the equations (3.1)-(3.2) hold at $\partial \Sigma_{c}$,
and the equations (3.4)-(3.5), or (3.6),
hold at $\partial \Sigma_{c}$.

Before imposing the boundary conditions (i), or (ii),
or (iii), the {\it constraint equations} (see previous
remarks) of complex general relativity read
(cf (2.2)-(2.3))
$$
\int_{\Sigma_{c}}\partial_{a}\sigma^{ab} \;
d^{3}x_{b}
-\int_{\Sigma_{c}}\lambda^{{\hat c}{\hat d}}
\Bigr(D_{a}{\tilde p}^{ab}\Bigr)_{{\hat c}{\hat d}} \;
d^{3}x_{b}=0
\eqno (4.1)
$$
$$
\int_{\Sigma_{c}}\partial_{a}\rho^{ab} \; d^{3}x_{b}
-\int_{\Sigma_{c}}{\rm Tr}
\biggr[{\tilde p}^{af}\Omega_{ad}-{1\over 2}{\tilde p}^{ab}
\Omega_{ab} \; \delta_{d}^{f}\biggr] u^{d} \;
d^{3}x_{f}=0.
\eqno (4.2)
$$
With our notation, $\Omega_{ab}^{\; \; \; {\hat c}{\hat d}}$
is the holomorphic curvature of the holomorphic connection 1-form
$\omega_{a}^{\; \; {\hat c}{\hat d}}$. Moreover, $D$ is a connection
which annihilates the internal-space metric $\eta_{{\hat a}{\hat b}}$
(cf [6]).
On imposing the boundary conditions studied so far, the first
term on the left-hand side of (4.1)-(4.2) vanishes, and the
preservation of constraints yields
the contracted Bianchi identities.
Thus, the full set of field equations linear in the holomorphic
multimomenta takes the form (cf [6])
$$
{\rm Tr} \Bigr[{\tilde p}^{ij}\Omega_{ij}\Bigr]=0
\eqno (4.3)
$$
$$
{\rm Tr} \Bigr[{\tilde p}^{i0}\Omega_{ij}\Bigr]=0
\eqno (4.4)
$$
$$
\Bigr(D_{a}{\tilde p}^{a0}\Bigr)_{{\hat c}{\hat d}}=0.
\eqno (4.5)
$$
We omit the details to avoid repeating the analysis appearing
in [6]. However, we should emphasize that (4.3)-(4.5) are
obtained by fixing the holomorphic coordinate $z^{0}$,
which does not have a distinguished role with respect to
$z^{1},z^{2},z^{3}$. Hence the interpretation of our
particular coordinate system is quite different from the
Lorentzian case. In other words,
the equations (4.3)-(4.5) {\it correspond}
to the Hamiltonian, momentum and Gauss constraints of
the Lorentzian theory, respectively, but they {\it should not}
be regarded as describing a 3+1 split of the
four-complex-dimensional geometry.

Note that it is not {\it a priori} obvious that the
three-complex-dimensional surface $\Sigma_{c}$ has no
boundary. Hence one really has to consider the boundary
conditions (ii) or (iii) in the holomorphic framework.
They imply that the holomorphic multimomenta have to
vanish everywhere on $\Sigma_{c}$ (by virtue of a
well-known result in complex analysis), or the elements
of $o(4,C)$ have to vanish everywhere on $\Sigma_{c}$,
jointly with the self-dual and anti-self-dual parts of the
connection 1-form. The latter of these conditions may be
replaced by the vanishing of the holomorphic vector field
$u$ on $\Sigma_{c}$. In other words, if $\Sigma_{c}$ has a
boundary, unless the holomorphic multimomenta vanish on
the whole of $\Sigma_{c}$, there are restrictions
at $\Sigma_{c}$ on the
spinor fields expressing the holomorphic nature of the
theory and its invariance under all holomorphic
coordinate transformations. Indeed, already in real Lorentzian
four-manifolds one faces a choice between boundary conditions on
the multimomenta and restrictions on the invariance group
resulting from boundary effects. We choose the former
and emphasize their role in
complex general relativity. Of course, the spinor fields
involved in the boundary conditions are instead non-vanishing
on the four-complex-dimensional space-time.

Remarkably, to ensure that the holomorphic multimomenta
${\tilde p}_{\; \; \;{\hat c}{\hat d}}^{ab}$ vanish at
$\partial \Sigma_{c}$, and hence on $\Sigma_{c}$ as well,
the determinant $e$ of the tetrad should vanish at
$\partial \Sigma_{c}$, or
$e^{-1} \; {\tilde p}_{\; \; \; {\hat c}{\hat d}}^{ab}$
should vanish at $\partial \Sigma_{c}$. The former case
admits as a subset the totally null two-complex-dimensional
surfaces known as $\alpha$-surfaces and $\beta$-surfaces
[1-2,5]. Since the integrability condition for
$\alpha$-surfaces is expressed by the vanishing of the
self-dual Weyl spinor, our formalism enables one to recover
the anti-self-dual (also called right-flat) space-time
relevant for twistor theory, where both the Ricci spinor
$R_{AA'BB'}$ and the self-dual
Weyl spinor ${\widetilde \psi}_{A'B'C'D'}$ vanish. However, if
$\partial \Sigma_{c}$ is not totally null, the resulting theory
does not correspond to twistor theory.
The latter case implies that the tetrad vectors are turned
into holomorphic vectors $u_{1},u_{2},u_{3},u_{4}$, say, such that
one of the following conditions holds
at $\partial \Sigma_{c}$, and hence on $\Sigma_{c}$ as well:
(i) $u_{1}=u_{2}=u_{3}=u_{4}=0$;
(ii) $u_{1}=u_{2}=u_{3}=0, u_{4} \not = 0$;
(iii) $u_{1}=u_{2}=0, u_{3}=\gamma u_{4}, \gamma \in {\cal C}$;
(iv) $u_{1}=0, \gamma_{2}u_{2}=\gamma_{3}u_{3}
=\gamma_{4}u_{4}, \gamma_{i} \in {\cal C}, i=2,3,4$;
(v) $\gamma_{1}u_{1}=\gamma_{2}u_{2}=\gamma_{3}u_{3}
=\gamma_{4}u_{4}, \gamma_{i} \in {\cal C}, i=1,2,3,4$.
\vskip 10cm
\leftline {\bf 5. Open problems}
\vskip 1cm
\noindent
It now appears essential to
understand the relation between complex general relativity
derived from jet-bundle theory and complex general
relativity as in the Penrose twistor programme.
For this purpose, one needs to study the topology and the
geometry of the space of two-complex-dimensional surfaces
$\partial \Sigma_{c}$ in the generic case. This leads to
a deep link between complex space-times which are not
anti-self-dual and two-complex-dimensional surfaces
which are not totally null. In other words, on going beyond
twistor theory, one finds that the analysis of
two-complex-dimensional surfaces still plays a key role.
Last, but not least, one has to solve equations
(cf (4.3)-(4.5)) which are now
linear in the {\it holomorphic multimomenta}, both in
classical and in quantum gravity (these equations correspond
to the constraint equations of the Lorentzian theory).
Hence we hope that our
paper may provide the first step towards a new
synthesis in relativistic theories of gravitation.
\vskip 1cm
\leftline {\bf Acknowledgments}
\vskip 1cm
\noindent
We are indebted to Giuseppe Marmo and Gabriele Gionti for
enlightening conversations on geometric methods in field
theory. Partial support by the European Union under the
Human Capital and Mobility Programme is gratefully acknowledged.
\vskip 10cm
\leftline {\bf Appendix}
\vskip 1cm
\noindent
To help the readers who are not familiar with multisymplectic
geometry, we present a very brief outline of jet bundles and
momentum maps.

In section 4, the notation $J^{1}$ means what follows. Let $X$
be a manifold, and let $Y$ be a fibre bundle having $X$ as
its base space, with projection map $\pi_{XY}$. Moreover, let
$\gamma : T_{x}X \rightarrow T_{y}Y$ be a linear map between
the tangent space to $X$ at $x$ and the tangent space to $Y$
at $y \in \pi_{XY}^{-1}(x)$. Given a point $y$ belonging to
the fibre $Y_{x}$ through $x \in X$, one considers all
$\gamma$ maps relative to $y \in Y_{x}$. This leads to a fibre
bundle $J^{1}(Y)$ having the fibre bundle $Y$ as its base space
and fibres given by the $\gamma$ maps. Such a $J^{1}(Y)$ is
called the one-jet bundle on $Y$.

A familiar property of classical mechanics and classical field
theory is that, if the Lagrangian is invariant under the action
of a group, by virtue of Noether's theorem there exist functions
which are constant along solutions of the equations of motion.
The constraints of a field theory result from Noether's theorem
through the action of the gauge group or the group of space-time
diffeomorphisms. The {\it covariant multimomentum map} is the
mathematical tool which enables one to describe these properties
of classical fields. In section 4, the covariant multimomentum
map ${\widetilde J}(u)$ reads
$$
{\widetilde J}(u) \equiv
\left[e {\partial L \over \partial
\omega_{a \; \; \; \; \; ,f}^{\; \; {\hat c}{\hat d}}}
\Bigr(u_{a}^{\; \; {\hat c}{\hat d}}
-\omega_{a \; \; \; \; ,b}^{\; \; {\hat c}{\hat d}}
\; u^{b}\Bigr)
+{e\over 2} e_{\; \; {\hat c}}^{a} \;
e_{\; \; {\hat d}}^{b} \;
\Omega_{ab}^{\; \; \; {\hat c}{\hat d}} \; u^{f}
\right] \; d^{3}x_{f}
\eqno (A1)
$$
where $L \equiv {e\over 2}e_{\; \; {\hat c}}^{a} \;
e_{\; \; {\hat d}}^{b} \;
\Omega_{ab}^{\; \; \; {\hat c}{\hat d}}$ is the Lagrangian.
With our notation, $u_{a}^{\; \; {\hat c}{\hat d}}$ describes
coordinate transformations along the fibre, and it is given
by
$$
u_{a}^{\; \; {\hat c}{\hat d}}=-u_{\; \; , a}^{b}
\; \omega_{b}^{\; \; {\hat c}{\hat d}}
+(D_{a}\lambda)^{{\hat c}{\hat d}}.
\eqno (A2)
$$
Moreover, the spinorial form of the holomorphic vector field
$u$ (see (3.6)) is obtained from the familiar relation
$u^{a} \; e_{a}^{\; \; AA'}=u^{AA'}$.
On integrating ${\widetilde J}(u)$ on $\Sigma_{c}$ and
setting such an integral to zero, the holomorphic
{\it constraint equations} (4.1)-(4.2) are obtained.
\vskip 1cm
\leftline {\bf References}
\vskip 0.3cm
\item {[1]}
Penrose R and Rindler W 1986 {\it Spinors and Space-Time II:
Spinor and Twistor Methods in Space-Time Geometry}
(Cambridge: Cambridge University Press)
\item {[2]}
Ward R S and Wells R O 1990 {\it Twistor Geometry and
Field Theory} (Cambridge: Cambridge University Press)
\item {[3]}
Ashtekar A 1991 {\it Lectures on Non-Perturbative
Canonical Gravity} (Singapore: World Scientific)
\item {[4]}
Esposito G 1994 {\it Quantum Gravity, Quantum Cosmology
and Lorentzian Geometries} (Lecture Notes in Physics {\bf m12})
(Berlin: Springer)
\item {[5]}
Esposito G 1995 {\it Complex General Relativity}
(Fundamental Theories of Physics {\bf 69})
(Dordrecht: Kluwer)
\item {[6]}
Esposito G, Gionti G and Stornaiolo C {\it Space-Time Covariant
Form of Ashtekar's Constraints} (DSF preprint 95/7)
\item {[7]}
Kaiser G 1981 {\it J. Math. Phys.} {\bf 22} 705
\bye